\documentclass[conference]{IEEEtran}
\IEEEoverridecommandlockouts

\usepackage{amsmath,amssymb,amsfonts}
\usepackage{graphicx}
\usepackage{textcomp}
\usepackage{hyperref}
\usepackage{titlesec}
\usepackage{dblfloatfix}  
\usepackage[protrusion=true,expansion=true,tracking=true]{microtype}
\usepackage{cite} 

\usepackage{booktabs}
\usepackage{pifont}
\usepackage{enumitem}

\usepackage{xspace}
\usepackage{color}
\usepackage{listings}
\usepackage{xcolor}

\definecolor{darkspringgreen}{rgb}{0.09, 0.45, 0.27}
\definecolor{denim}{rgb}{0.08, 0.38, 0.74}
\definecolor{darkolivegreen}{rgb}{0.33, 0.42, 0.18}
\definecolor{tangerine}{rgb}{0.95, 0.52, 0.0}
\definecolor{mahogany}{rgb}{0.75, 0.25, 0.0}

\definecolor{uglyyellow}{rgb}{0.99, 0.93, 0.0}
\definecolor{nbs}{rgb}{0.35, 0.31, 0.81}

\definecolor{darkred}{rgb}{0.75, 0.1, 0.1}

    \newcommand{\vnit}[1]{\textcolor{black}{#1}}                

\definecolor{seagreen}{rgb}{0.18, 0.55, 0.34}

\definecolor{darkpink}{rgb}{0.88, 0.28, 0.54}
\definecolor{forestgreen}{rgb}{0.0, 0.27, 0.13}
\definecolor{amber}{rgb}{1.0, 0.49, 0.0}

\newcommand{\head}[1]{{\noindent\textbf{#1.}\xspace}} 
\newcommand{\inum}[1]{(\textit{#1})\xspace}

\newcommand{\RMV}{\textsf{RMV}\xspace}            
\newcommand{\PMV}{\textsf{PMV}\xspace}            

\usepackage{tikz}
\usetikzlibrary{positioning,arrows.meta,shapes.geometric,calc,fit,backgrounds}

\newcommand{\squishlist}{
 \begin{list}{$\circ$}
  { \setlength{\itemsep}{0pt}
     \setlength{\parsep}{0pt}
     \setlength{\topsep}{3pt}
     \setlength{\partopsep}{0pt}
     \setlength{\leftmargin}{1em}
     \setlength{\labelwidth}{1em}
     \setlength{\labelsep}{0.5em} } }

\newcommand{\squishend}{
  \end{list}  }

\newcommand*\circled[1]{\tikz[baseline=(char.base)]{\node[shape=circle,fill,inner sep=0.5pt] (char) {\textcolor{white}{#1}};}}

\definecolor{denim}{rgb}{0.08, 0.38, 0.74}
\definecolor{azure(colorwheel)}{rgb}{0.0, 0.5, 1.0}
\definecolor{greenp}{rgb}{0.0, 0.65, 0.50}
\definecolor{peach}{rgb}{0.97, 0.51, 0.47}
\definecolor{darkmagenta}{rgb}{0.55, 0.0, 0.55}
\definecolor{royalblue(web)}{rgb}{0.25, 0.41, 0.88}
\definecolor{ao(english)}{rgb}{0.0, 0.5, 0.0}
\definecolor{ForestGreen}{RGB}{34,139,34}

\usepackage[colorinlistoftodos,prependcaption,textsize=scriptsize]{todonotes}

\usepackage[bottom]{footmisc}
\usepackage{dblfloatfix}
\usepackage{array}
 \usepackage{booktabs}

\definecolor{ufogreen}{rgb}{0.1, 0.6, 0.4}
\definecolor{ufogreen}{rgb}{0.1, 0.6, 0.4}

\newcommand{\system}{Argus}

\usepackage{listings}
\usepackage{xcolor}
\usepackage[T1]{fontenc} 
\usepackage{zi4} 
\usepackage{caption} 
\usepackage{float} 

\lstdefinestyle{custompseudocode}{
  belowcaptionskip=1\baselineskip,
  breaklines=true,
  xleftmargin=\parindent,
  language=Python,
  showstringspaces=false,
  basicstyle=\small\ttfamily,
  keywordstyle=\bfseries\color{green!40!black},
  commentstyle=\itshape\color{purple},
  stringstyle=\color{orange},
  numbers=left,
  numberstyle=\scriptsize\color{black},
  numbersep=8pt,
  morekeywords={function}, 
  keywordstyle=[2]\bfseries, 
  escapeinside={*@}{@*}, 
  xleftmargin=2em, xrightmargin=0em, 
}

\setlist{nosep,leftmargin=*}

\makeatletter
\renewenvironment{abstract}
  {\par\noindent\normalfont\normalsize\upshape\mdseries
   \textbf{Abstract.}\enspace\ignorespaces}
  {\par}
\makeatother

\renewcommand{\thesection}{\arabic{section}}
\renewcommand{\thesubsection}{\thesection.\arabic{subsection}}
\renewcommand{\thesubsubsection}{\thesubsection.\arabic{subsubsection}}

\titleformat{\section}{\normalfont\large\bfseries}{\thesection}{0.6em}{}
\titleformat{\subsection}{\normalfont\normalsize\bfseries}{\thesubsection}{0.5em}{}
\titleformat{\subsubsection}{\normalfont\normalsize\bfseries\itshape}{\thesubsubsection}{0.5em}{}
\titlespacing*{\section}{0pt}{6pt}{2pt}
\titlespacing*{\subsection}{0pt}{4pt}{1pt}
\titlespacing*{\subsubsection}{0pt}{3pt}{1pt}

\begin{document}

\bstctlcite{argus:bibstyle}

\title{\system{}: \underline{A}gentic, \underline{R}eference-Calibrated,
  Tree-\underline{G}\underline{u}ided, \underline{S}ystem-Software-Level
  Bottleneck Localization\\
}

\author{%
\IEEEauthorblockN{Vlad-Petru Nitu \quad Harsh Songara \quad Konstantinos Sgouras\\
Spiros Galanopoulos \quad Konstantinos Kanellopoulos \quad Onur Mutlu}
\IEEEauthorblockA{ETH Z\"urich}}

\maketitle

\begin{abstract}
Operating system (OS) code can account for a substantial share of CPU execution time.
\emph{First}, as application logic is
offloaded to heterogeneous accelerators (e.g., GPUs), the CPU
increasingly acts as an orchestrator, spending cycles in driver calls,
data movement, and synchronization rather than in application code.
\emph{Second}, workloads such as serverless functions frequently invoke OS services.
At the same time, the OS is a complex codebase
spanning many subsystems (e.g., memory management, networking), making
it hard to localize the specific code path responsible for a slowdown.
Existing profilers expose measurements that require interpretation (e.g., \textit{perf} and \textit{Intel VTune}) or can perturb short operations when extensively instrumented (e.g., \textit{ftrace}).
Diagnosing OS bottlenecks can therefore require repeated kernel instrumentation and manual interpretation.

We introduce \textbf{\system{}}, 
an agentic LLM-based profiler that produces
instrumentation code and autonomously (i.e., without human intervention) reasons over potential OS-level bottlenecks.
\system{} integrates two key mechanisms: (i)~a \emph{calibration} methodology that involves
collecting a measurement from an idle system and using it as a reference point to discover potential bottlenecks, 
and (ii)~a \emph{tree-based data structure} that represents the different OS execution paths, improving the agent's bottleneck localization accuracy.
\system{} aims to identify a specific kernel code path rather than stop at a subsystem-level diagnosis.
In two case studies, we employ \system{} to autonomously discover bottlenecks present in the memory management subsystem caused by
(i) a THP aggressor co-running with other applications,
and (ii) applications that incur different types of page faults.
\system{} produces \vnit{\textbf{$19\times$}} fewer incorrect deep-path diagnoses than the strongest evaluated LLM-based baseline, which lacks reference calibration,
while preserving low time-to-diagnosis ($\approx 31s$).
\end{abstract}


\section{Introduction}
\label{sec:intro}

OS routines can be substantial performance bottlenecks.
In a study of warehouse-scale workloads, Kanev et al. report that
kernel code accounts for nearly one fifth of CPU cycles, with scheduling
alone representing more than 5\%~\cite{kanev15datacentertax}.
Virtual-memory operations such as page faults and physical-page allocation can also impose substantial software costs~\cite{kanellopoulos25virtuoso}.
Two workload trends make OS diagnosis particularly important.
First, accelerators (e.g., GPUs and TPUs~\cite{hennessy19goldenage,jouppi17tpu}) still rely on host software; GPU drivers, for example, manage device memory, interrupts, and synchronization~\cite{amdgpu}. GPU page-size and demand-paging choices can also affect application performance~\cite{ausavarungnirun17mosaic}.
Second, workloads involving microsecond-scale I/O~\cite{barroso2017killermicroseconds,cho18killerus}, virtualization~\cite{agache20firecracker}, and microservices~\cite{gan19deathstarbench} often invoke OS services repeatedly, so delays in those services can increase end-to-end request latency.

\head{What existing profilers miss}
We group representative profilers into six classes.
(1)~Time-based samplers (e.g., \texttt{perf} using software clock events~\cite{perf}) reveal \emph{where} CPU time is spent, but not \emph{why}.
(2)~Hardware-event samplers (e.g., \texttt{perf} using PMU events~\cite{perf,vtune,uprof,yasin2014tma}) cannot directly measure how long a thread waits for OS services.
(3)~Static tracepoints expose predefined kernel events~\cite{perf,desnoyers2006lttng,kernel-tracepoints}; their overhead rises with event frequency and the work done per event (e.g., recording a stack trace).
(4)~Continuous profilers (e.g., Grafana Pyroscope~\cite{pyroscope,parca,datadog_profiler})
collect CPU samples, which cannot measure blocking time (e.g., how long a thread waits to acquire a kernel mutex)~\cite{pyroscope-profile-types,datadog-profile-types}.
(5)~Accelerator profilers include NVIDIA Nsight Systems, which traces GPU activity and host-device calls~\cite{nsight_systems}, and AMD Omniperf, which reports GPU-kernel metrics~\cite{omniperf}. Nsight Systems can also sample host-kernel stacks~\cite{nsight-user-guide}. \system{} uses its reference profile and kernel tree to select finer probes.
(6)~Dynamic instrumentation systems (e.g., DTrace and SystemTap~\cite{cantrill2004dtrace,prasad2005systemtap})
can attach probes without recompilation. Linux kprobes supports most kernel instruction addresses~\cite{kprobes-doc}.
However, writing these hooks requires kernel expertise, severely limiting the adoption of such tools. 

Extended Berkeley Packet Filter (eBPF) programs~\cite{gbadamosi24ebpf} enable programmable kernel instrumentation at supported hooks and serve as the basis for tools such as
bpftrace~\cite{bpftrace} and eunomia~\cite{eunomia}.
eBPF is backed by a
verifier~\cite{ebpf-verifier} that statically checks each program for
properties such as bounded execution and valid memory access before loading it into the kernel.
Once loaded, eBPF code can aggregate measurements in kernel maps, reducing the volume of data copied to user space.
We quantify eBPF's overhead against other profilers (e.g., \texttt{ftrace}) in \S\ref{sec:evaluation}.

However, flexible kernel instrumentation has two key limitations.
First, it requires manual authoring of complex eBPF code, which demands familiarity with kernel internals, making it unsuitable for developers without deep kernel expertise. We call this the \emph{authoring gap}.
Second, these profiling tools still leave the interpretation of results and the choice of follow-up probes to the user. We call this the \emph{interpretation gap}.
This gap arises because bottleneck localization is an iterative process: starting from coarse subsystem-level signals, the engineer progressively narrows the search to the specific kernel execution path responsible for the slowdown.
Each iteration demands expertise: interpreting the current
measurement, instrumenting the next finer-grained probe, and
re-profiling.
BIS and GAPP identify serialization bottlenecks in multithreaded applications~\cite{joao12bis,gapp20}; Haecki et al. diagnose end-host network latency~\cite{haecki22nanos}. These methods do not use a kernel code-path hierarchy and an idle-system reference to choose which functions to probe next.
Ideally, a diagnosis tool should reach the kernel execution path
responsible for the bottleneck, rather than stop at the broader
subsystem.

\head{LLM-based agents create a new opportunity}
Recent LLM-based agents~\cite{opencode,alphaevolve}
\footnote{An
\emph{LLM-based agent} is a large language model (LLM) that interacts with
external tools in a loop: it observes outputs, decides on an action,
executes it, and repeats until it completes the task.}
can author code and use tool feedback to reason about a task,
creating an opportunity to close both the authoring and
interpretation gaps. However, naively using agents for system-level profiling has three limitations.
\emph{First}, an agent needs a reference value: only a significant deviation of the observed measurement from this reference indicates a bottleneck.
\emph{Second}, such reference values are noisy and vary across
hardware configurations.
\emph{Third}, without a structured view of the kernel's code paths,
the agent invents function names or guesses an incorrect fine-grained diagnosis.
We quantify the impact of all three limitations in \S\ref{sec:evaluation}.

\textbf{Our goal} in this work is to develop an autonomous OS-level profiling methodology that is low-overhead, programmable, and supports dynamic instrumentation, while closing the \emph{authoring} and \emph{interpretation} gaps.
To this end, we propose \textbf{\system{}}, an agentic profiler that
\emph{integrates} an LLM-based agent with two complementary mechanisms:
\emph{reference calibration} against an idle-system profile, and
a \emph{tree-based data structure} that represents the different OS execution paths.
\system{} descends the tree one level at a time, comparing each level's
measurements against the idle-system reference to decide whether to
commit to the flagged code path or descend deeper.

Across a (victim, perturber) matrix of five workloads and four perturbations,
\system{} produces $19\times$ fewer incorrect deep-path diagnoses than the strongest evaluated baseline (i.e., \system{} without reference calibration),
at comparable time-to-diagnosis ($\approx31s$ vs. $\approx29s$).

In this work, we make the following contributions:
(i)~we propose \system\xspace, an autonomous OS-level
profiler that employs an LLM agent to localize bottlenecks in OS code;
(ii)~we compare three baselines that omit reference calibration and/or tree-guided traversal, observing incorrect-diagnosis rates $19$--$31\times$ higher than \system{}'s;
and (iii)~we use \system{} to autonomously discover bottlenecks in two memory-management case studies.

\section{\system{}: Key Idea, Overview, and Mechanism}
\label{sec:overview}

We propose \system{}, an agentic profiler that autonomously
diagnoses system-level performance bottlenecks down to the responsible OS
code path.
The \textbf{key idea} of \system{} is to
\emph{guide} the LLM agent via two complementary mechanisms:
(i)~\emph{reference calibration}, which compares the agent's
measurements against a baseline profile of the workload collected on
an idle reference system,
and (ii)~\emph{tree-guided kernel
traversal}, which constrains the agent's exploration to a static,
hierarchical tree of kernel code paths built offline.
By doing so, \system{} validates bottleneck hypotheses under
perturbation while avoiding hallucinated, non-existent code paths.
This enables autonomous bottleneck localization at code-path
granularity, without expert intervention.

\begin{figure}[!ht]
\centering
\includegraphics[width=\columnwidth,keepaspectratio]{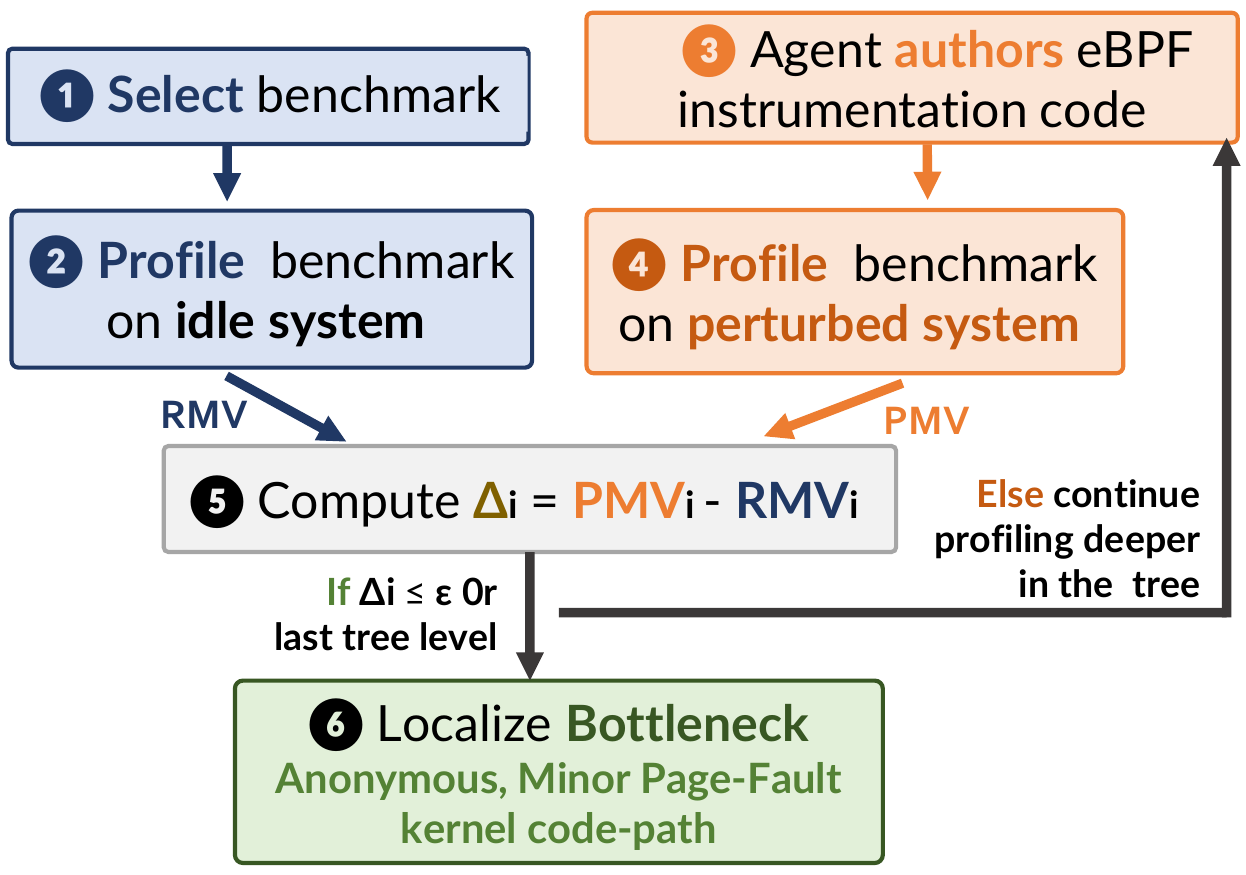}
\caption{Overview of \system{}'s key components.}
\label{fig:argus-pipeline}
\end{figure}

Fig.~\ref{fig:argus-pipeline} illustrates \system{}'s high-level
components. The reference and perturbed measurement vectors (RMV/PMV)
each capture perf counters and per-code-path latencies.
\system{} operates in six steps.
\circled{1}~The user selects a workload $M$ to profile.
\circled{2}~\system{} profiles $M$ on an idle system using \texttt{perf}~\cite{perf} and eBPF probes,
yielding the \RMV{}.
\circled{3}~The agent authors a single \emph{multiplexed} eBPF probe
(i.e., one program that hooks every branch of the tree at the current
level $L$).
\circled{4}~\system{} profiles $M$ on the perturbed system using the loaded eBPF probe,
yielding the \PMV{}.
\circled{5}~\system{} computes $\Delta_i=\mathrm{PMV}_i-\mathrm{RMV}_i$ for each measured kernel-path metric $i$.
$\Delta$ is used to compare \PMV{} against \RMV{} element-wise to decide whether to terminate the search (i.e., no bottleneck found), or descend one level deeper into the flagged subtree (i.e., significant deviation in at least one dimension of the vector),  in which case it returns to step~\circled{3}.
\circled{6}~Upon termination, \system{} reports the final diagnosis:
the flagged kernel code path,
or \emph{no bottleneck found},
if \system{} abstained
(i.e., no measured value in the \PMV{} deviates significantly from its reference in the \RMV{}).

\begin{figure}[!ht]
\centering
\includegraphics[width=\columnwidth]{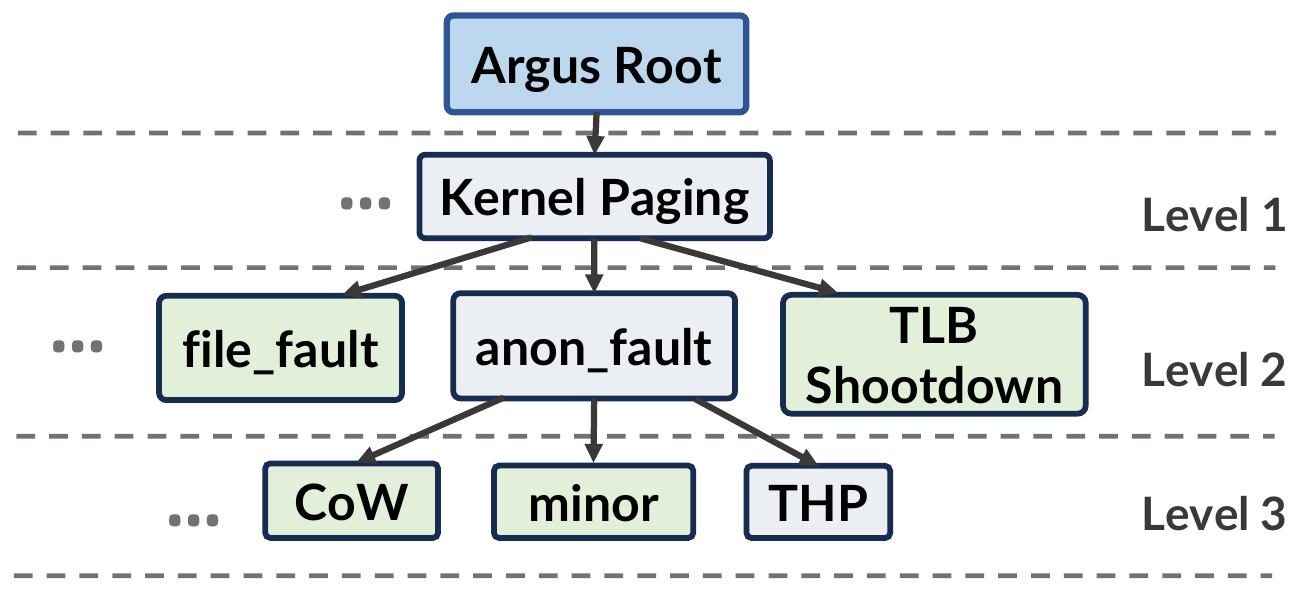}
\caption{The paging branch of \system{}'s three-level kernel code-path tree.}
\label{fig:argus-tree}
\end{figure}

\head{Hierarchical subsystem tree} Fig.~\ref{fig:argus-tree} shows the paging branch of \system{}'s static kernel code-path tree.
Level~1 identifies a kernel subsystem (e.g., kernel paging).
Level~2
breaks it into paths such as \texttt{file\_fault}, \texttt{anon\_fault},
and TLB shootdown. Level~3 refines \texttt{anon\_fault} into CoW, minor,
and THP paths.
The tree bounds this traversal to three levels, with one multiplexed probe per level.

\head{Reference profiling~\circled{2}} \system{} produces the
\RMV{} by profiling workload $M$ on an idle reference system, which serves as
the baseline against which the agent later detects deviations under
perturbation. \system{} profiles $M$ under \texttt{perf}~\cite{perf}
and multiplexed eBPF probes to populate the
\RMV{} with per-code-path call counts and latency numbers
(e.g., \texttt{do\_anonymous\_page}).

\head{Agent profiling closed loop~\circled{3}-\circled{4}} At each level $L$, the
agent performs three steps to produce the \PMV{}.
\inum{i}
\emph{Generate}: the agent generates eBPF source code that hooks the
level-$L$ code paths in the active subtree (i.e., the subtree rooted at
the path flagged at level $L\!-\!1$, or the entire tree at $L\!=\!1$).
At $L\!=\!2$ and $L\!=\!3$, the agent multiplexes all hooks into a
single eBPF program that shares counters through one eBPF
map~\cite{eBPF-shared-maps}, such that one program covers the entire
subtree.
\inum{ii} \emph{Verify and load}: the agent submits the
generated program to the eBPF verifier. If the verifier accepts the
program, \system{} loads it into the kernel. Otherwise, the agent
re-prompts itself with the verifier's diagnostic message in a
ReAct loop~\cite{yao23react}, and repeats step~\inum{i} until the
program is accepted.
\inum{iii} \emph{Measure}: \system{} runs $M$ on
the perturbed system,
and collects the counters sampled using the code generated in step (i) into the \PMV{}.

\head{Bottleneck detection and diagnosis reporting~\circled{5}-\circled{6}}
For each dimension $i$, \system{} computes
$\Delta_i = (\mathrm{PMV}_i - \mathrm{RMV}_i)$
and flags dimensions with $\Delta_i
> \tau$ (the \emph{confidence condition}; we set $\tau = 3$
empirically). If no dimension is flagged, \system{} reports \emph{no
bottleneck found}. If at least one dimension is flagged and $L$ is a
leaf, \system{} reports the flagged path as the bottleneck
(e.g., \texttt{kernel\_paging $\to$ anon\_fault $\to$ minor}).
If at least one dimension is flagged but $L$ is not a leaf, \system{} descends one level into the flagged child subtree and returns to step~\circled{3}.
By combining reference calibration with tree guidance, \system{} traverses at most three tree levels before reporting or abstaining.

\section{Evaluation}
\label{sec:evaluation}

To isolate the contribution of each component, we define three
baselines. (i)~\emph{B-LLM-Prior}: the agent receives no reference
profile and descends solely based on LLM's prior knowledge.
This is the weakest
baseline we evaluate, and isolates the contribution of using LLMs' intuition naively in an agentic profiler.
(ii)~\emph{B-LLM-Prior-Tree}: \emph{B-LLM-Prior} augmented with a static tree data structure that reflects a subset of the Linux source code,
isolating the contribution of tree-guided traversal.
(iii)~\emph{B-LLM-Prior-Tree-Probes}: augments \emph{B-LLM-Prior-Tree} by using an LLM-based agent to write eBPF code at every level of the tree,
which guides bottleneck localization with measured counters.
Unlike \system{}, this baseline omits reference calibration.
This baseline isolates the benefit of guiding tree traversal using measured counters.

\begin{figure}[!t]
\centering
\includegraphics[width=\columnwidth,keepaspectratio]{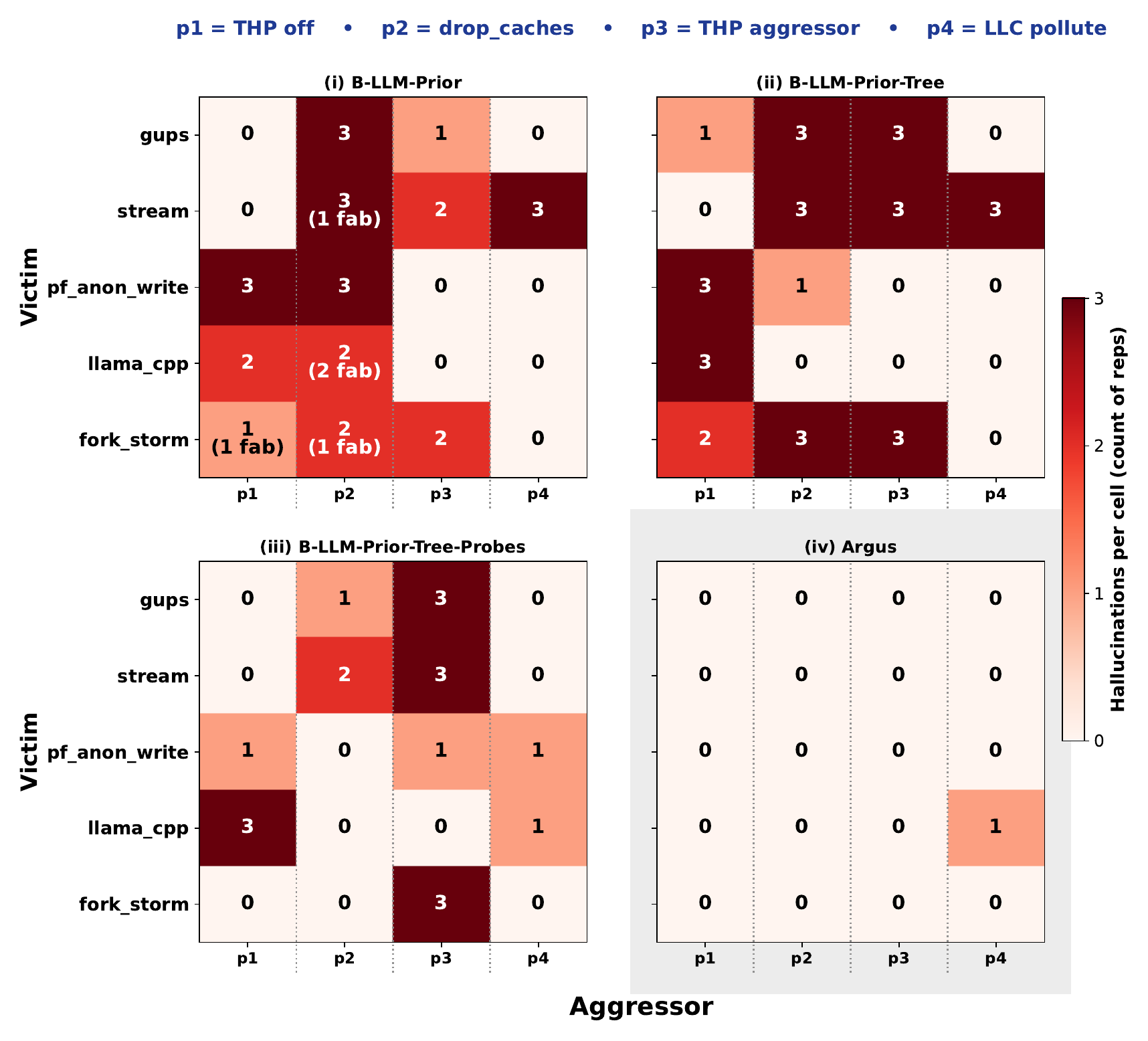}
\caption{Per-cell hallucinated diagnoses across (victim, perturber)
pairs, for \system{} and three LLM baselines.}
\label{fig:hallucation_per_cell}
\end{figure}

\head{Accuracy}
We construct a (victim, perturber)
matrix that pairs five representative workloads with four
perturbations (e.g., disable THP), and run each
baseline three times per cell.
We define a \textbf{hallucination} as a bottleneck diagnosis at a deep code path that does not match the cell's true bottleneck.
Hallucinations are either
\textbf{wrong attributions} (real kernel functions the workload does not exercise),
or \textbf{fabrications} (function names absent from the kernel, e.g., \texttt{alloc\_cold}).
Fig.~\ref{fig:hallucation_per_cell} reports the number of hallucinated diagnoses per (victim, perturber) cell for each baseline.
We make four key observations. 
\emph{First}, B-LLM-Prior hallucinates even on simple microbenchmarks because it bases its decisions on the perturber's name rather than the victim's actual access pattern.
\emph{Second}, B-LLM-Prior \emph{fabricates} kernel function
names (e.g., \texttt{alloc\_cold}, \texttt{page\_alloc}),
since it lacks a structured view of the kernel's code paths,
which \system{} provides through the tree-based data structure.
\emph{Third}, without a reference measurement, B-LLM-Prior has no notion of \emph{deviation} and cannot recognize null cells (i.e., no bottleneck),
sometimes committing deeply in the tree with little confidence.
In contrast, \system{}
correctly abstains when the perturbation has no measurable effect. \\
\emph{Fourth}, providing the tree-based data structure associated with the kernel (B-LLM-Prior-Tree, B-LLM-Prior-Tree-Probes) removes fabricated
hallucinations entirely, since the tree restricts the bottleneck localization to existing kernel functions.
However, B-LLM-Prior-Tree still descends deep without enough evidence, sometimes hallucinating more than B-LLM-Prior. \\
B-LLM-Prior-Tree-Probes outperforms the other two LLM-based baselines,
but it does not match \system{}: without a reference measurement, its probes cannot establish whether a measured value deviates from normal operation.
We conclude that enhancing the agentic traversal with a tree that mirrors the kernel, without a reference profile, is
insufficient to achieve high bottleneck localization accuracy, motivating the \emph{synergistic interplay of
reference calibration and tree-guidance}.

\begin{figure}[t]
\centering
\includegraphics[width=\columnwidth,keepaspectratio]{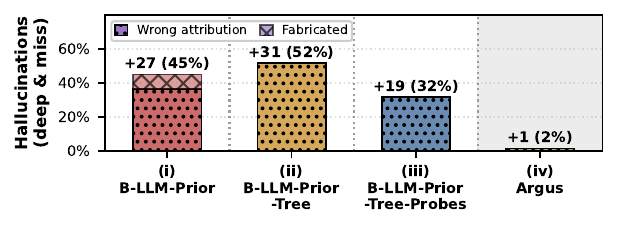}
\caption{Incorrect deep-path diagnosis rate across all (victim,
perturber) pairs. \system{}'s rate is \vnit{$19$--$31\times$} lower than
the evaluated LLM-based baselines.}
\label{fig:hallucation}
\end{figure}

Fig.~\ref{fig:hallucation} aggregates these results across all (victim, perturber) pairs.
We make two key observations.
\emph{First}, \system{} makes an incorrect deep-path diagnosis in \vnit{one of 60 runs ($1.7\%$)}, compared
to \vnit{27 of 60 runs ($45\%$)} for B-LLM-Prior and \vnit{31 of 60 runs ($51.7\%$)} for B-LLM-Prior-Tree
(\vnit{$27\times$ and $31\times$ higher rates, respectively}).
B-LLM-Prior-Tree-Probes makes incorrect diagnoses in \vnit{19 of 60 runs ($31.7\%$)}. \vnit{Compared with this
best-performing baseline, \system{} has a $19\times$ lower incorrect-diagnosis rate.}
\emph{Second}, while B-LLM-Prior terminates exploration when uncertain, B-LLM-Prior-Tree descends the tree without sufficient evidence, misidentifying the code path on which the bottleneck lies.
These results show that, in this matrix, adding a tree and probes without reference calibration does not prevent incorrect deep-path diagnoses.

\begin{figure}[!ht]
\centering
\includegraphics[width=\columnwidth,keepaspectratio]{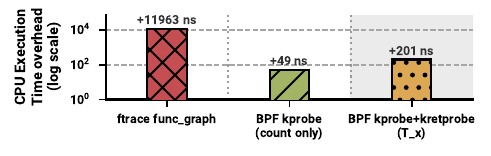}
\caption{CPU execution time overhead (log scale) of three
kernel-tracing options over an uninstrumented baseline:
\texttt{ftrace}'s \texttt{func\_graph}, an eBPF \texttt{kprobe} that
only counts events, and an eBPF \texttt{kprobe}+\texttt{kretprobe}
that captures per-call latency $T_x$.}
\label{fig:overhead-cpu}
\end{figure}

\head{Overhead analysis}
Fig.~\ref{fig:overhead-cpu} quantifies the CPU execution time overhead
per kernel function call on a fault-heavy microbenchmark
(\texttt{pf\_anon\_write} $\times$ \texttt{p1}, which sustains
$2.6$\,M \texttt{do\_anonymous\_page} calls over $3$\,s).
We make three key observations.
\emph{First},
\texttt{ftrace}'s~\cite{ftrace} \texttt{func\_graph} tracer emits
entry and exit events for each traced function call, incurring \vnit{$11.9\,\mu$s} per call in this setup.
\emph{Second}, a count-only eBPF \texttt{kprobe} reduces this overhead
by \vnit{$243\times$}, to \vnit{$49$\,ns} per call, but loses per-call latency
information. \emph{Third}, an eBPF \texttt{kprobe}+\texttt{kretprobe}
pair that captures per-call entry/exit latency $T_x$ in-kernel incurs
\vnit{$201$\,ns} per call, at only \vnit{$4\times$} more than the count-only
probe, but \vnit{$60\times$} less than \texttt{ftrace}. We conclude that
the \texttt{kprobe}+\texttt{kretprobe} configuration represents the sweet spot
between low overhead and diagnostic signal: at a fraction of
\texttt{ftrace}'s cost, $T_x$ provides the agent with the per-call
latency it needs to attribute slowdowns to specific kernel code paths.
\system{} therefore adopts this configuration.

\head{Time-to-Diagnosis} We evaluate the
median time-to-diagnosis (TtD) of \system{} and the three baselines
defined above, on the same (victim, perturber) matrix
used in the hallucination study (Fig.~\ref{fig:hallucation_per_cell}).
We make two key observations.
\emph{First}, \system{} has a TtD in the same range as the LLM-prior baselines:
its \vnit{$30.9$\,s}
median is on par with all three (B-LLM-Prior: \vnit{$33.4$\,s}, B-LLM-Prior-Tree: \vnit{$26.3$\,s}, B-LLM-Prior-Tree-Probes: \vnit{$28.7$\,s}),
despite the additional reference profiling and reference calibration steps that \system{} performs.
\emph{Second}, B-LLM-Prior is \vnit{$17\%$} slower than the
average of the other three configurations,
because the agent oscillates
between subsystem hypotheses without a tree to guide its traversal.
In this evaluation, reference calibration adds little to median TtD while reducing incorrect deep-path diagnoses.

\section{Case Studies}
\label{sec:workflows}
We demonstrate \system{} in two case studies: (1)~\vnit{per-victim
diagnosis under a THP aggressor}, and (2)~disambiguating which code
paths different page-fault types stress within the paging module of the 
memory management subsystem.

\begin{figure}[!ht]
\centering
\includegraphics[width=\columnwidth,keepaspectratio]{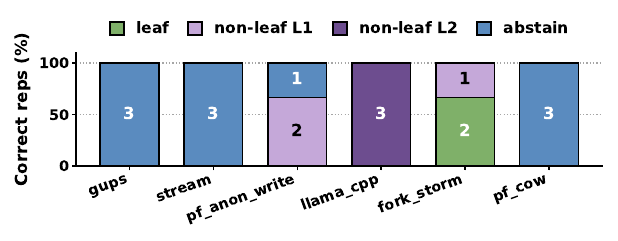}
\caption{\system{}'s diagnosis outcomes for six victim workloads
co-running with \texttt{p10\_thp\_aggressor}. Each bar shows three
repetitions, grouped by diagnosis depth or abstention; the numbers
on the bars represent repetition counts.}
\label{fig:cs1-thp-aggressor}
\end{figure}

\head{Case Study~1: THP aggressor}
Fig.~\ref{fig:cs1-thp-aggressor} shows \system{}'s outcomes for six
victim workloads co-running with \texttt{p10\_thp\_aggressor}. This
workload fragments physical memory and then forces compaction by
requesting huge pages. Prior work also examines huge-page performance
under memory fragmentation~\cite{kanellopoulos26revelator}.
\emph{First}, \textbf{gups} and \textbf{stream} (3/3 abstain)
pre-allocate in userspace, so the aggressor never reaches their hot
path. Thus, \system{} abstains for both workloads in all three repetitions.
\emph{Second}, \textbf{pf\_anon\_write} (3/3 correct) commits to
\texttt{kernel\_paging} at L1 in 2/3 reps.
\vnit{\emph{Third}, \textbf{llama\_cpp} (3/3 correct) exposes a
non-obvious kernel-paging signal: although GPU-bound, its memory-mapped
model file generates \texttt{file\_fault} activity under the
aggressor's page-cache pressure, while KV-cache growth drives
\texttt{anon\_fault}. \system{} commits to \texttt{kernel\_paging} at
L2.}
\emph{Fourth}, for \textbf{fork\_storm}, \system{} correctly commits to
\texttt{kernel\_scheduler.load\_balance} at leaf level 2/3 times.
\emph{Fifth},
\textbf{pf\_cow} (3/3 abstain) exposes a tree limitation: the THP tax
spreads across \texttt{mmap\_lock}, slab, and RCU paths that the
kernel tree of code paths does not contain. \\
The missing paths in the current tree limit the diagnosis for
\textbf{pf\_cow}. We leave expanding the tree (e.g., auto-generating it) as future work.

\begin{figure}[!t]
\centering
\includegraphics[width=\columnwidth,keepaspectratio]{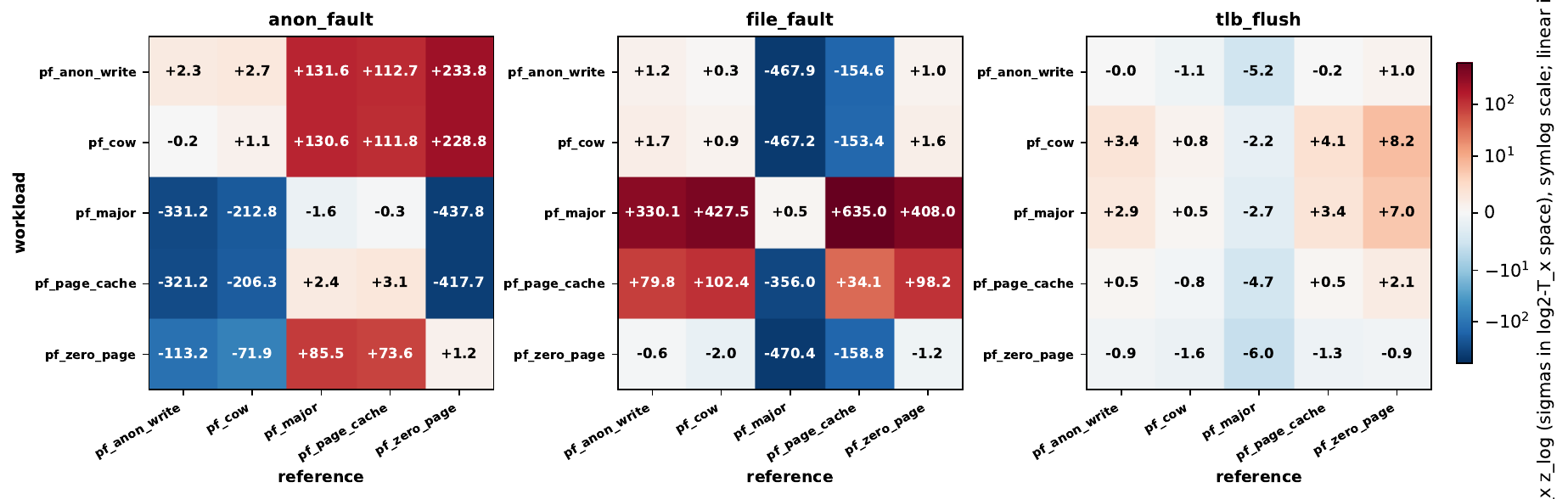}
\caption{Page-fault disambiguation: $T_x$ signatures of five page-fault
perturbers across three kernel code paths: \texttt{anon\_fault},
\texttt{file\_fault}, and \texttt{tlb\_flush}.}
\label{fig:cs2-pf-l2}
\end{figure}

\head{Case Study~2: Page-fault disambiguation}
Fig.~\ref{fig:cs2-pf-l2} reports the $T_x$ measured by \system{} for
five page-fault perturbers across three kernel code paths:
\texttt{anon\_fault}, \texttt{file\_fault}, and \texttt{tlb\_flush}.
The \texttt{pf\_cow} workload exercises Copy-on-Write, whose page copying can add substantial memory-management cost~\cite{seshadri15pageoverlays}.
Three clusters emerge. \emph{First}, the \texttt{anon\_fault} cluster
(\texttt{pf\_anon\_write}, \texttt{pf\_cow}, \texttt{pf\_zero\_page})
shows high $T_x$ on the anonymous path and near-zero $T_x$ on the
file-backed path, confirming that anonymous and file-backed workloads
exercise disjoint kernel paths. \emph{Second}, in \texttt{file\_fault},
\texttt{pf\_major} exhibits ${\sim}5\times$ higher $T_x$ than
\texttt{pf\_page\_cache}, separating ``file-backed fault'' from
``file-backed fault \emph{with disk I/O}.'' \emph{Third}, the
\texttt{tlb\_flush} cluster (\texttt{pf\_cow}, \texttt{pf\_major})
shows lower $T_x$, since TLB invalidations are events
that fire only on Copy-on-Write and swap-in remapping.

We conclude that the five perturbers resolve into four signature
classes: (1)~\emph{anon-only} (\texttt{pf\_anon\_write},
\texttt{pf\_zero\_page}); (2)~\emph{anon+TLB} (\texttt{pf\_cow});
(3)~\emph{file-backed-light} (\texttt{pf\_page\_cache}); and
(4)~\emph{file-backed-with-IO} (\texttt{pf\_major}). \system{} recovers
the kernel's page-fault taxonomy without prior knowledge
about the perturber.

\bibliographystyle{SAFARItran}
\bibliography{references}

@misc{perf,
  author = {{Linux Kernel}},
  title  = {\texttt{perf}: Linux profiling with performance counters},
  howpublished = {\url{https://perfwiki.github.io/main/}},
  note   = {Accessed: 2026-04-24}
}

@misc{vtune,
  author = {{Intel Corporation}},
  title  = {Intel VTune Profiler},
  howpublished = {\url{https://www.intel.com/content/www/us/en/developer/tools/oneapi/vtune-profiler.html}},
  note   = {Accessed: 2026-04-24}
}

@misc{uprof,
  author = {{Advanced Micro Devices}},
  title  = {AMD {\textmu{}Prof} Profiler},
  howpublished = {\url{https://www.amd.com/en/developer/uprof.html}},
  note   = {Accessed: 2026-04-24}
}

@inproceedings{yasin2014tma,
  author    = {Yasin, Ahmad},
  title     = {A {Top-Down} Method for Performance Analysis and Counters Architecture},
  booktitle = {ISPASS},
  year      = {2014},
  pages     = {35--44},
  doi       = {10.1109/ISPASS.2014.6844459},
}

@misc{ftrace,
  author = {Rostedt, Steven},
  title  = {\texttt{ftrace}: Function Tracer for the Linux Kernel},
  howpublished = {\url{https://www.kernel.org/doc/html/latest/trace/ftrace.html}},
  note   = {Accessed: 2026-04-24}
}

@misc{kprobes-doc,
  author = {Keniston, Jim and Panchamukhi, Prasanna S. and Hiramatsu, Masami},
  title = {Kernel Probes ({Kprobes})},
  howpublished = {\url{https://docs.kernel.org/trace/kprobes.html}},
  note = {Accessed: 2026-09-28}
}

@misc{kernel-tracepoints,
  author = {Desnoyers, Mathieu},
  title = {Using the {Linux} Kernel Tracepoints},
  howpublished = {\url{https://docs.kernel.org/trace/tracepoints.html}},
  note = {Accessed: 2026-09-28}
}

@inproceedings{desnoyers2006lttng,
  author    = {Desnoyers, Mathieu and Dagenais, Michel R.},
  title     = {The {LTTng} tracer: A low impact performance and behavior monitor for {GNU/Linux}},
  booktitle = {OLS},
  year      = {2006},
  pages     = {209--224},
}

@inproceedings{cantrill2004dtrace,
  author    = {Cantrill, Bryan M. and Shapiro, Michael W. and Leventhal, Adam H.},
  title     = {Dynamic Instrumentation of Production Systems},
  booktitle = {USENIX ATC},
  year      = {2004},
  pages     = {15--28},
}

@inproceedings{prasad2005systemtap,
  author    = {Prasad, V. and Cohen, W. and Eigler, F.C. and Hunt, M. and Keniston, J. and Chen, B.},
  title     = {Locating System Problems Using Dynamic Instrumentation},
  booktitle = {OLS},
  year      = {2005},
  pages     = {49--64},
}

@misc{pyroscope,
  author = {{Grafana Labs}},
  title  = {Grafana Pyroscope: Continuous Profiling},
  howpublished = {\url{https://grafana.com/oss/pyroscope/}},
  note   = {Accessed: 2026-04-24}
}

@misc{pyroscope-profile-types,
  author = {{Grafana Labs}},
  title = {Profiling Types and Their Uses},
  howpublished = {\url{https://grafana.com/docs/pyroscope/latest/introduction/profiling-types/}},
  note = {Accessed: 2026-09-28}
}

@misc{datadog-profile-types,
  author = {{Datadog}},
  title = {Profile Types},
  howpublished = {\url{https://docs.datadoghq.com/profiler/profile_types/}},
  note = {Accessed: 2026-09-28}
}

@misc{parca,
  author = {{Polar Signals}},
  title  = {Parca: Continuous Profiling for Analysis of CPU and Memory Usage},
  howpublished = {\url{https://www.parca.dev/}},
  note   = {Accessed: 2026-04-24}
}

@misc{datadog_profiler,
  author = {{Datadog}},
  title  = {Datadog Continuous Profiler},
  howpublished = {\url{https://docs.datadoghq.com/profiler/}},
  note   = {Accessed: 2026-04-24}
}

@misc{nsight_systems,
  author = {{NVIDIA Corporation}},
  title  = {{NVIDIA} {Nsight} {Systems}},
  howpublished = {\url{https://developer.nvidia.com/nsight-systems}},
  note   = {Accessed: 2026-04-25}
}

@misc{nsight-user-guide,
  author = {{NVIDIA Corporation}},
  title = {{Nsight Systems} User Guide},
  howpublished = {\url{https://docs.nvidia.com/nsight-systems/UserGuide/}},
  note = {Accessed: 2026-09-28}
}

@misc{omniperf,
  author = {{Advanced Micro Devices}},
  title  = {{Omniperf} Documentation},
  howpublished = {\url{https://rocm.docs.amd.com/projects/omniperf/en/docs-6.2.4/}},
  note   = {Accessed: 2026-04-25}
}

@misc{bpftrace,
  author = {{bpftrace Project}},
  title  = {\texttt{bpftrace}: High-level Tracing Language for {Linux}},
  howpublished = {\url{https://github.com/bpftrace/bpftrace}},
  note   = {Accessed: 2026-04-24}
}

@misc{eunomia,
  author = {{eunomia-bpf Project}},
  title  = {eunomia-bpf: A Dynamic Loading Library and Compile Toolchain for {eBPF}},
  howpublished = {\url{https://github.com/eunomia-bpf/eunomia-bpf}},
  note   = {Accessed: 2026-04-24}
}

@inproceedings{gapp20,
  author    = {Reena Nair and Tony Field},
  title     = {{GAPP}: A Fast Profiler for Detecting Serialization Bottlenecks in Parallel {Linux} Applications},
  booktitle = {ICPE},
  year      = {2020},
  pages     = {257--264},
  doi       = {10.1145/3358960.3379136},
  note      = {arXiv:2004.05628}
}

@inproceedings{haecki22nanos,
  author    = {Roni Haecki and Radhika Niranjan Mysore and Lalith Suresh and Gerd Zellweger and Bo Gan and Timothy Merrifield and Sujata Banerjee and Timothy Roscoe},
  title     = {How to Diagnose Nanosecond Network Latencies in Rich End-host Stacks},
  booktitle = {NSDI},
  year      = {2022},
  pages     = {861--877},
}

@inproceedings{cho18killerus,
  author    = {Shenghsun Cho and Amoghavarsha Suresh and Tapti Palit and Michael Ferdman and Nima Honarmand},
  title     = {Taming the Killer Microsecond},
  booktitle = {MICRO},
  year      = {2018},
  pages     = {627--640},
  doi       = {10.1109/MICRO.2018.00057}
}

@article{barroso2017killermicroseconds,
  author    = {Luiz Andr{\'e} Barroso and Mike Marty and David Patterson and Parthasarathy Ranganathan},
  title     = {Attack of the Killer Microseconds},
  journal   = {Communications of the ACM},
  volume    = {60},
  number    = {4},
  pages     = {48--54},
  year      = {2017},
  doi       = {10.1145/3015146}
}

@inproceedings{kanev15datacentertax,
  author    = {Kanev, Svilen and Darago, Juan Pablo and
               Hazelwood, Kim and Ranganathan, Parthasarathy and
               Moseley, Tipp and Wei, Gu-Yeon and Brooks, David},
  title     = {Profiling a Warehouse-Scale Computer},
  booktitle = {ISCA},
  year      = {2015},
  pages     = {158--169},
  doi       = {10.1145/2749469.2750392}
}

@misc{amdgpu,
  author = {{Linux Kernel Contributors}},
  title  = {{AMDGPU}: {Linux} Kernel Driver for {AMD} {GPU}s},
  howpublished = {\url{https://docs.kernel.org/gpu/amdgpu/index.html}},
  note   = {Accessed: 2026-05-07}
}

@inproceedings{yao23react,
  author    = {Yao, Shunyu and Zhao, Jeffrey and Yu, Dian and
               Du, Nan and Shafran, Izhak and Narasimhan, Karthik and
               Cao, Yuan},
  title     = {{ReAct}: Synergizing Reasoning and Acting in Language Models},
  booktitle = {ICLR},
  year      = {2023},
  note      = {arXiv:2210.03629}
}

@misc{eBPF-shared-maps,
  author = {{ebpf.io Authors}},
  title  = {{eBPF} Maps},
  howpublished = {\url{https://docs.ebpf.io/linux/concepts/maps/}},
  note   = {Accessed: 2026-05-07}
}

@misc{ebpf-verifier,
  author = {{Linux Kernel Contributors}},
  title  = {{eBPF} verifier},
  howpublished = {\url{https://docs.kernel.org/bpf/verifier.html}},
  note   = {Accessed: 2026-05-07}
}

@misc{opencode,
  author = {{Anomaly}},
  title  = {{OpenCode}: The open source {AI} coding agent},
  howpublished = {\url{https://opencode.ai}},
  note   = {Source repository: \url{https://github.com/anomalyco/opencode}.
            Accessed: 2026-05-07}
}

@misc{alphaevolve,
  author = {Novikov, Alexander and V{\~u}, Ng{\^a}n and
            Eisenberger, Marvin and Dupont, Emilien and
            Huang, Po-Sen and Wagner, Adam Zsolt and
            Shirobokov, Sergey and Kozlovskii, Borislav and
            Ruiz, Francisco J. R. and Mehrabian, Abbas and
            Kumar, M. Pawan and See, Abigail and
            Chaudhuri, Swarat and Holland, George and
            Davies, Alex and Nowozin, Sebastian and
            Kohli, Pushmeet and Balog, Matej},
  title  = {{AlphaEvolve}: A Coding Agent for Scientific
            and Algorithmic Discovery},
  year   = {2025},
  howpublished = {Google {DeepMind} White Paper.
                  \url{https://arxiv.org/abs/2506.13131}},
  note   = {arXiv:2506.13131}
}

@inproceedings{gan19deathstarbench,
  author    = {Gan, Yu and Zhang, Yanqi and Cheng, Dailun and Shetty, Ankitha and Rathi, Priyal and others},
  title     = {An Open-Source Benchmark Suite for Microservices and Their Hardware-Software Implications for Cloud \& Edge Systems},
  booktitle = {ASPLOS},
  year      = {2019},
  pages     = {3--18},
  doi       = {10.1145/3297858.3304013}
}

@inproceedings{agache20firecracker,
  author    = {Agache, Alexandru and Brooker, Marc and Florescu, Andreea and Iordache, Alexandra and Liguori, Anthony and Neugebauer, Rolf and Piwonka, Phil and Popa, Diana-Maria},
  title     = {Firecracker: Lightweight Virtualization for Serverless Applications},
  booktitle = {NSDI},
  year      = {2020},
  pages     = {419--434}
}

@article{hennessy19goldenage,
  author  = {Hennessy, John L. and Patterson, David A.},
  title   = {A New Golden Age for Computer Architecture},
  journal = {Communications of the ACM},
  volume  = {62},
  number  = {2},
  pages   = {48--60},
  year    = {2019},
  doi     = {10.1145/3282307}
}

@inproceedings{jouppi17tpu,
  title     = {In-Datacenter Performance Analysis of a Tensor Processing Unit},
  author    = {Jouppi, Norman P. and Young, Cliff and Patil, Nishant and Patterson, David and Agrawal, Gaurav and Bajwa, Raminder and Bates, Sarah and Bhatia, Suresh and Boden, Nan and Borchers, Al and others},
  booktitle = {ISCA},
  pages     = {1--12},
  year      = {2017},
  doi       = {10.1145/3079856.3080246}
}

@misc{gbadamosi24ebpf,
  title         = {The {eBPF} Runtime in the {Linux} Kernel},
  author        = {Gbadamosi, Bolaji and Leonardi, Luigi and Pulls, Tobias and H{\o}iland-J{\o}rgensen, Toke and Ferlin-Reiter, Simone and Sorce, Simo and Brunstr{\"o}m, Anna},
  year          = {2024},
  eprint        = {2410.00026},
  archivePrefix = {arXiv},
  primaryClass  = {cs.OS},
  url           = {https://arxiv.org/abs/2410.00026}
}

@inproceedings{kanellopoulos25virtuoso,
  author    = {Konstantinos Kanellopoulos and Konstantinos Sgouras and F. Nisa Bostanci and Andreas Kosmas Kakolyris and Berkin Kerim Konar and Rahul Bera and Mohammad Sadrosadati and Rakesh Kumar and Nandita Vijaykumar and Onur Mutlu},
  title     = {{Virtuoso}: Enabling Fast and Accurate Virtual Memory Research via an Imitation-based Operating System Simulation Methodology},
  booktitle = {ASPLOS},
  year      = {2025},
  pages     = {1400--1421},
  doi       = {10.1145/3676641.3716027}
}

@inproceedings{ausavarungnirun17mosaic,
  author    = {Rachata Ausavarungnirun and Joshua Landgraf and Vance Miller and Saugata Ghose and Jayneel Gandhi and Christopher J. Rossbach and Onur Mutlu},
  title     = {{Mosaic}: A {GPU} Memory Manager with Application-Transparent Support for Multiple Page Sizes},
  booktitle = {MICRO},
  year      = {2017},
  pages     = {136--150},
  doi       = {10.1145/3123939.3123975}
}

@inproceedings{joao12bis,
  author    = {Jos{\'e} A. Joao and M. Aater Suleman and Onur Mutlu and Yale N. Patt},
  title     = {Bottleneck Identification and Scheduling in Multithreaded Applications},
  booktitle = {ASPLOS},
  year      = {2012},
  pages     = {223--234},
  doi       = {10.1145/2150976.2151001}
}

@inproceedings{kanellopoulos26revelator,
  author    = {Konstantinos Kanellopoulos and Konstantinos Sgouras and Harsh Songara and Andreas Kosmas Kakolyris and Vlad-Petru Nitu and Spiros Galanopoulos and Rahul Bera and Konstantina Koliogeorgi and Rakesh Kumar and Onur Mutlu},
  title     = {{Revelator}: Rapid Data Fetching via System-Software-Guided Hash-based Speculative Address Translation},
  booktitle = {ISCA},
  year      = {2026},
  pages     = {2061--2080},
  doi       = {10.1109/ISCA66397.2026.00147}
}

@inproceedings{seshadri15pageoverlays,
  author    = {Vivek Seshadri and Gennady Pekhimenko and Olatunji Ruwase and Onur Mutlu and Phillip B. Gibbons and Michael A. Kozuch and Todd C. Mowry and Trishul Chilimbi},
  title     = {Page Overlays: An Enhanced Virtual Memory Framework to Enable Fine-grained Memory Management},
  booktitle = {ISCA},
  year      = {2015},
  pages     = {79--91},
  doi       = {10.1145/2749469.2750379}
}

@IEEEtranBSTCTL{argus:bibstyle,
  CTLuse_forced_etal       = {yes},
  CTLmax_names_forced_etal = {2},
  CTLnames_show_etal       = {1}
}


\end{document}